\documentclass[journal]{IEEEtran}
\usepackage{url}
\usepackage{xcolor}
\usepackage{tikz}
\usetikzlibrary{arrows.meta,positioning,fit}

\usepackage{cite}
\usepackage{amsmath,amssymb,amsfonts}
\usepackage{algorithmic}
\usepackage{graphicx}
\usepackage{subcaption}
\usepackage{booktabs}
\usepackage{multirow}
\usepackage{array}

\usepackage{amsthm}
\theoremstyle{remark}

\usepackage{xcolor}
\usepackage{textcomp}
\def\BibTeX{{\rm B\kern-.05em{\sc i\kern-.025em b}\kern-.08em
		T\kern-.1667em\lower.7ex\hbox{E}\kern-.125emX}}
\usepackage{nomencl}
\makenomenclature
\usepackage{etoolbox}

\usepackage{fancyhdr}

\fancypagestyle{plain}{
  \fancyhf{} 
  \fancyhead[L]{\small\textcolor{blue}{This work has been accepted by IEEE Transactions on Industrial Informatics on 2026-09-09. Digital Object Identifier (DOI) by IEEE Transactions on Industrial Informatics: 10.1109/TII.2026.3734758.}}
}
\begin{document}

	\title{Hierarchical Graph Neural Network for Probabilistic Power Flow with\\ Grid Topology Contingencies}
	\author{Jinbao Wang, Jun Liu, Shiliang Zhang, Xuehui Ma, Lin Yan, Haolin Liu, Sabita Maharjan\thanks{Jinbao Wang, Jun Liu, Xuehui Ma, Lin Yan, Haolin Liu are with School of Automation and Information Engineering, Xi'an University of Technology, No.5 Jinhua South Road, Xi'an, 710048, Shaanxi, China. Shiliang Zhang (shilianz@ifi.uio.no) and Sabita Maharjan (sabita@ifi.uio.no) are with the Department of Informatics, University of Oslo, 0373 Oslo, Norway. \\ \textcolor{blue}{© 2026 IEEE.  Personal use of this material is permitted.  Permission from IEEE must be obtained for all other uses, in any current or future media, including reprinting/republishing this material for advertising or promotional purposes, creating new collective works, for resale or redistribution to servers or lists, or reuse of any copyrighted component of this work in other works.}} }
	
	\maketitle
    \thispagestyle{plain}
	
	\begin{abstract}
		Data-driven surrogates reduce the nonlinear solves required for probabilistic power flow (PPF), but models tied to a fixed bus--branch graph often generalize poorly when outages change electrical coupling. This paper proposes a topology-decomposition-driven hierarchical bidirectional graph neural network (TDD--HBGNN) for PPF under unseen topologies. For each realized topology, the network is decomposed into fundamental topological units (FTUs), dynamically homogeneous voltage clusters (DHVCs), and a core backbone sub-grid (CBSG), with equivalent-admittance representations preserving electrical ports between units. A sensitivity-based linear prior captures the first-order voltage-magnitude and phase-angle response, while the HBGNN learns topology-dependent nonlinear residuals. Impedance- and uncertainty-aware bottom-up aggregation condenses local perturbations, while mode-modulated top-down broadcasting restores system context at buses. We evaluate our approach on the IEEE 118-bus and Polish 2746-bus systems under unseen N--1 and N--2 contingencies against six baseline methods. Compared with these methods, TDD--HBGNN achieves 2.47--4.20-fold lower all-bus phase-angle errors and lower phase-angle errors in every held-out event. Its voltage-magnitude advantage is concentrated near the outaged branches, where paired error ratios reach 1.75--2.58 on the IEEE system and 1.87--1.97 in Polish N--2 events. All system--contingency--baseline comparisons favor TDD--HBGNN in joint voltage--angle error. Moreover, shared hierarchical operators keep the model size at 0.658 million parameters across both grids and yield 0.161--1.793~ms forward latency per probabilistic case. These results demonstrate system-wide angular accuracy, outage-local voltage accuracy, and parameter-efficient scaling across unseen topologies.
	\end{abstract}
	
	
	\mbox{}
	
	\nomenclature[A]{$\mathcal{G}_{bone}$}{CBSG subgraph}
	\nomenclature[A]{$l$}{Hierarchical level index, $l \in \{0,1,2,3\}$}
	\nomenclature[A]{$K$}{Number of bidirectional message-passing rounds}
	\nomenclature[A]{$M$}{Number of mode components}
	\nomenclature[A]{$d$}{Hidden feature dimension}
	\nomenclature[A]{$n_c$}{Number of DHVC clusters}
	\nomenclature[A]{$r$}{Number of retained principal components}
	\nomenclature[A]{$N_s$}{Number of historical time snapshots}

	\nomenclature[B]{$S_{PV}$}{Sensitivity matrix $\partial V / \partial P$}
	\nomenclature[B]{$S_{P\theta}$}{Sensitivity matrix $\partial \theta / \partial P$}
	\nomenclature[B]{$C_{ij}^{G}$}{Granger causality strength from $i$ to $j$}
	\nomenclature[B]{$\mathcal{H}_\delta$}{Huber loss with threshold $\delta$}
	\nomenclature[B]{$\boldsymbol{\Phi}$}{Regularized regression coefficient matrix for residual voltage extraction}
	\nomenclature[B]{$A_{ij}$}{Affinity matrix entry between nodes $i$ and $j$}
	\nomenclature[B]{$\mathbf{Y}_{eq}$}{Equivalent admittance matrix of a topological unit}

	\nomenclature[C]{$\mathbf{e}_{ij}$}{Edge feature vector $[r_{ij}, x_{ij}, |z_{ij}|] \in \mathbb{R}^3$}
	\nomenclature[C]{$\epsilon_i$}{Local uncertainty scalar at node $i$}
	\nomenclature[C]{$\mathbf{h}_i^{(l)}$}{Feature vector of node $i$ at level $l$}
	\nomenclature[C]{$\mathbf{z}_i$}{Mode affinity vector of node $i$, $\mathbf{z}_i \in \mathbb{R}^M$}
	\nomenclature[C]{$\mathbf{G}_{ij}$}{Edge-level gating vector from node $j$ to $i$, $\in \mathbb{R}^d$}
	\nomenclature[C]{$\boldsymbol{\gamma}_i$}{Node-level gating vector for top-down fusion, $\in \mathbb{R}^d$}
	\nomenclature[C]{$\alpha_j$}{Normalized $\epsilon$-aware attention weight for topology-to-cluster pooling}
	\nomenclature[C]{$\beta_{bi}$}{Normalized attention weight from backbone node $b$ to node $i$}
	\nomenclature[C]{$Y_{eq,ij}$}{Equivalent admittance of edge $(i,j)$}
	\nomenclature[C]{$\mathbf{m}_{ij}$}{Message from source node $j$ to target node $i$}
	\nomenclature[C]{$\mathcal{V}^{(l)}$}{Vertex set at hierarchical level $l$}
	\nomenclature[C]{$\pi^{(l)}$}{Membership mapping from level $l{-}1$ to level $l$}
	\nomenclature[C]{$\mathcal{N}_i$}{Set of neighbors of node $i$}
	\nomenclature[C]{$\mathcal{C}_p, \mathcal{C}_c, \mathcal{C}_{bo}$}{Node set of topology $p$; topology set of cluster $c$; cluster set of backbone $bo$}
	\nomenclature[C]{$\ell(e)$}{Edge labeling function}
	\nomenclature[C]{$\ell_B(\mathcal{G}_B)$}{Block classification function}
	\nomenclature[C]{$\mathcal{G}_R$}{Residual graph after removing transformer and switch edges}
	\nomenclature[C]{$\operatorname{int}(\cdot)$}{Interior vertex set}

\nomenclature[D]{$\mathrm{MLP}_{(\cdot)}^{(\cdot)}(\cdot)$}{Multi-layer perceptron with ReLU activation; subscript denotes function (e.g., $g$: gate, $m$: message, $fuse$: fusion, $\alpha$: attention, $base$: center quantile, $q^{\pm}$: upper/lower increment); superscript denotes inter-level mapping ($^{(l-1 \to l)}$: bottom-up; $^{(l \to l-1)}$: top-down)}
\nomenclature[D]{$\sigma(\cdot)$}{Sigmoid activation function}
\nomenclature[D]{$\odot$}{Hadamard (element-wise) product}

\nomenclature[S]{$Z_{ij}$}{Electrical impedance distance between node $i$ and $j$}
\nomenclature[S]{$\mathbf{W}_{(\cdot)}$}{Learnable weight matrix; subscript denotes function ($m$: message, $u$: update, $y$: admittance projection, $g$: fusion gate, $mp$: mode projection, $bc^{mode}$: backbone broadcast, $mode$: mode modulation, $attn$: attention)}
\nomenclature[S]{$[\cdot \| \cdot]$}{Vector concatenation}

\printnomenclature[4em]
\section{Introduction}
\IEEEPARstart{A}{s} the energy transition rapidly progresses, uncertainty in power-system operation grows because of the large-scale integration of distributed energy resources and the intermittency of renewable generation. Probabilistic power flow (PPF) analysis has therefore emerged as a critical tool for evaluating the effects of these uncertainties. Rooted in probability theory, PPF aims to obtain the statistical characteristics of bus-voltage magnitudes and phase angles. It operates by modeling nodal power injections as random variables and solving the resulting stochastic power-flow mappings over the grid structure \cite{2023Probabilistic}. However, traditional PPF methods require extensive repeated power-flow calculations \cite{2022Statistical,gao2023physics}. Consequently, their computational burden increases as the system scale, number of uncertainty scenarios, and dimensionality of the stochastic variables grow. This burden limits the use of PPF in online analysis and rapid decision-making for large-scale grids.

Data-driven methods have recently emerged as feasible tools for constructing fast power-flow approximations with nonlinear representation capacity. Deep surrogates replace repeated nonlinear solves with learned mappings from uncertain injections to system states or power-flow forecasts, substantially reducing the online cost of probabilistic and predictive analysis \cite{xiao2023deepppf,2024Bayesian,beinert2023forecast}. Physics-informed learning further regularizes these mappings with grid equations or operational constraints \cite{huang2023applications,nellikkath2022pinn}. Computational efficiency and physical regularization, however, do not by themselves ensure structural generalization. A topology-agnostic network encodes electrical coupling only through its training distribution; when a branch outage or reconfiguration changes the adjacency and admittance relations, the physical mapping changes without a corresponding change in the model's computation. Graph-based surrogates address this mismatch by coupling their propagation to the realized network structure \cite{gao2023physics,liu2023topologyaware,yang2024topologytransferable,yaniv2023adoption,dejongh2022geometric}. Yet accepting a different graph is not equivalent to generalizing to an unseen topology. Because the number of configurations grows combinatorially with the number of branches, exhaustive training remains infeasible, and successful transfer requires a model to preserve the electrical propagation mechanisms shared across configurations while adapting their interaction paths to the current grid.

Recent GNN-based power-flow models can be organized by the mechanism used to represent these interactions. Typed and topology-aware models expose component roles, line attributes, switch states, or the realized graph operator to message passing, thereby improving adaptation to heterogeneous or varying configurations \cite{lopezgarcia2023typed,liu2023topologyaware,yang2024topologytransferable,lopezgarcia2025typedopf,authier2024graphyr,gao2024physicsguidedopf}. Model-driven and physics-guided models instead derive propagation or training constraints from admittance relations, sensitivities, branch-current models, and power-flow equations, which improves physical consistency and reduces the burden on purely empirical fitting \cite{gao2023physics,yang2024ppf,ngo2024graphicalse}. High-order filters and recurrent or decentralized graph solvers enlarge the effective interaction range beyond immediate neighbors while retaining localized computation \cite{chen2023learning,lin2024powerflownet,donon2020neural,hansen2022power}, whereas multi-fidelity residual learning uses a low-fidelity power-flow approximation to reduce the demand for high-fidelity training data \cite{taghizadeh2024multifidelity}. Beyond solver learning, related power-system GNNs augment local propagation with higher-order node features, explicit network measures, or system-, zone-, and branch-level outputs to recover broader operating context \cite{chen2024real,zhang2025operationalrisk,zhu2025network}. These mechanisms address component heterogeneity, topology dependence, physical feasibility, propagation range, data efficiency, and system-level awareness. Most, however, still organize computation at a single bus--branch resolution. Repeated or high-order message passing can extend the receptive field, and graph-level summaries can inject global information, but neither explicitly separates contingency-local effects, regional aggregation, and system-wide feedback. Thus, the unresolved issue is not simply to propagate farther or impose more physics; it is to represent how a local topological perturbation is transformed across physically meaningful spatial scales.

This distinction also motivates the comparison and contradiction among the existing methods. Along the structural axis, a DNN provides a topology-agnostic global mapping, whereas a standard GNN, PowerFlowNet, and GNS progressively introduce bus-level graph propagation, higher-order aggregation, and iterative state updates \cite{xiao2023deepppf,lopezgarcia2023typed,lin2024powerflownet,donon2020neural}. Along the physical axis, the MD-GCN and PG-GNN incorporate model-derived coupling or power-flow constraints into graph learning \cite{gao2023physics,yang2024ppf}. This comparison separates the effects of graph structure, propagation range, iterative refinement, and embedded physics. None of these alternatives, however, explicitly reconstructs a changed grid as nested physical units or exchanges information bidirectionally between local units, regional aggregates, and the system backbone. The multiscale structural mechanism by which topology changes reshape stochastic power-flow mappings remains insufficiently investigated.

In summary, although existing data-driven PPF and power-flow models can fit grids with known or represented topologies, they do not explicitly characterize how topological perturbations systematically alter uncertainty-propagation paths and inter-nodal coupling strengths \cite{zhang2026power}. There remains a gap in representing the structural principles that govern these changes. Specifically, two critical research questions remain: (i) how do topological changes reshape stochastic power-flow mappings across local, regional, and system scales, and (ii) how can a structurally robust PPF propagation framework combine the dominant physical response with topology-dependent nonlinear corrections? These questions require a model whose propagation structure changes with the grid while preserving the physical mechanisms shared across configurations.

This paper addresses the insufficient topological generalization and poor structural adaptability of existing PPF models. To this end, we present a novel topology-decomposition-driven hierarchical bidirectional graph neural network (TDD--HBGNN) that sequentially decomposes the power grid into core backbone sub-grids (CBSGs), dynamically homogeneous voltage clusters (DHVCs), and fundamental topological units (FTUs). This decomposition follows fundamental electrical laws and is constrained by the physical grid structure. A sensitivity-based linearization establishes a physical baseline for the dominant power-flow response. The HBGNN then represents the grid through two complementary paths: (i) a bottom-up path that aggregates node features layer by layer through the FTUs and DHVCs to the CBSG, and (ii) a top-down path that broadcasts global contextual information back to individual nodes. Finally, the linearized baseline and the residual correction learned by the HBGNN jointly determine the bus-wise voltage-magnitude and phase-angle quantiles. Because the hierarchy and its propagation paths are reconstructed for the current grid, the model can generalize to previously unseen topological configurations. The main contributions are summarized as follows:
\begin{itemize}
	\item[1.] We propose an electrically constrained topology decomposition that partitions each grid configuration into FTUs, DHVCs, and a CBSG. It combines nodal electrical features with connectivity to form a physically meaningful multiscale representation of changing grid topologies.
	\item[2.] We develop TDD--HBGNN, which aggregates information from nodes through FTUs and DHVCs to the CBSG, then broadcasts system-level context to nodes. Reconstructing the hierarchy for each configuration couples local topology changes with global conditions under previously unseen topologies.
  
	\item[3.] We formulate PPF prediction as a sensitivity-based prior plus a topology-dependent nonlinear residual. The prior captures the dominant first-order response, while the HBGNN learns the residual to produce bus-wise voltage-magnitude and phase-angle quantiles.

\end{itemize}

\section{Problem formulation and framework of the proposed method}\label{motivation}
In this section, we formulate the problem in PPF, motivate the use of the topology-decomposition-driven HBGNN (TDD-HBGNN), and describe its architecture. We develop PPF based on the polar coordinate formulation of alternating current (AC) power flow.
\begin{equation}
	\tilde s = f(x),
\end{equation}
where $x$ denotes the vector of voltage angles and magnitudes at all buses except the slack bus, $\tilde s$ represents the vector of active and reactive nodal power injections, treated as random variables, and $f$  denotes the AC power flow mapping.

We employ model-independent quantiles to represent the uncertain power injection. The quantile set for a power grid with $q$ uncertain injection points at time step $t$ takes the form
\begin{equation}
	\mathcal{Q}_t = \bigcup_{i=1}^{q} \mathcal{Q}_{i,t}
	= \bigcup_{i=1}^{q}
	\left\{ P_{i,t,j} \right\}_{j=1}^{L},
\end{equation}
where $\mathcal{Q}_t$ denotes the overall quantile set at time $t$, $\mathcal{Q}_{i,t}$ is the quantile set for the $i$-th injection point, $P_{i,t,j}$ is the $j$-th quantile value of the uncertain power at the $i$-th injection point, and $L$ is the number of quantiles per injection point. The nominal proportions $\tau_{j}$ for each $P_{i,t,j}$ satisfy the standard quantile ordering
\begin{equation}
	0 < \tau_{1} < \tau_{2} < \cdots < \tau_{L} < 1.
\end{equation}

The model operates per time step. Note that we omit the index $t$ hereafter for brevity. The probabilistic outputs for voltage magnitude $\hat{V}$ and phase angle $\hat{\theta}$ at each bus follow separate quantile grids with $L_V$ and $L_\theta$ levels, respectively, where both are subsets of the $L$ input quantile levels.

\begin{figure}[!htb]
	\centering
	\captionsetup[subfigure]{font=scriptsize,labelfont=scriptsize}
	\subfloat[\label{Fig:DD_suba}]{\includegraphics[width=0.7\linewidth]{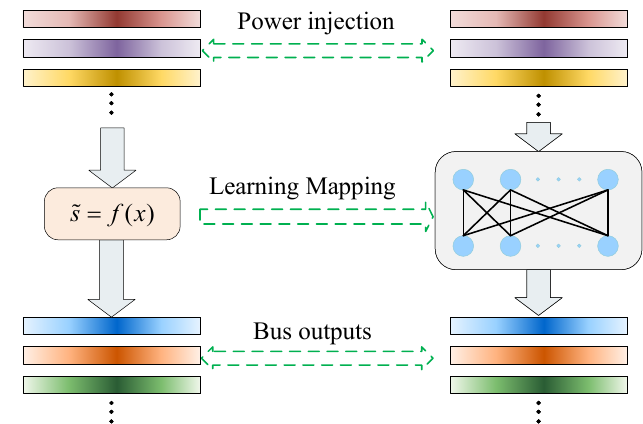}}
	\vspace{0.5cm} %
	\subfloat[\label{Fig:DD_subb}]{\includegraphics[width=0.7\linewidth]{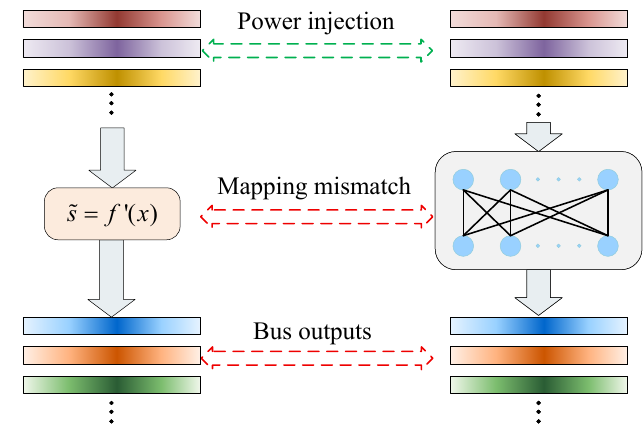}}
	\caption{Challenges of traditional data-driven methods. (a) The data-driven framework. The left side indicates the model-driven approach. The right side indicates the data-driven method learning the distribution mapping from data. (b) The data-driven framework challenged by grid topology mismatch. In this case, a new model $\tilde s = f'(x)$ is adopted to fit the grid, whereas the data-driven branch cannot adapt to the changed mapping, resulting in reduced accuracy.}\label{Fig:data_driven_a}
\end{figure}

Existing data-driven methods use historical or simulated data to learn the mapping from power injection distributions to voltage states, as illustrated in Fig. \ref{Fig:DD_suba}. However, when the network topology changes, the resulting mismatch between training and real-time data causes a marked decline in model performance, as shown in Fig. \ref{Fig:DD_subb}. This limitation underscores the need to improve the topological adaptability of these methods and to strengthen their ability to generalize to out-of-distribution scenarios.

\section{The proposed method}\label{method}
The structural properties of power systems can be examined along two dimensions: electrical dynamics in the grid and topological configuration. From an operational standpoint, a power system represents an integrated system driven by a CBSG and multiple DHVCs. From a topological standpoint, it forms a complex network built upon FTUs such as series, parallel, and mesh connections. Guided by this dual perspective, we develop HBGNN to enable multi-scale information propagation across hierarchical levels, thereby fully exploiting both the intrinsic electrical dynamics and the topology of the power system.

\subsection{Grid topological decomposition}\label{topological_decomposition}
By integrating network topology with historical input-output data, we first partition the entire system into a CBSG and multiple DHVCs. On this basis, we further decompose these components into minimal FTUs and linearize each unit, establishing a rigorous foundation for subsequent graph reconstruction and model training.

Let $\mathcal{G} = (\mathcal{V}, \mathcal{E})$ denote the power grid graph, where $\mathcal{V}$ and $\mathcal{E}$ represent the vertex and edge sets, respectively. For any subgraph $\mathcal{G}' = (\mathcal{V}', \mathcal{E}')$, we write $\deg_{\mathcal{G}'}(v)$ for the degree of vertex $v$ within $\mathcal{G}'$, and $\partial e = \{u, v\}$ for the pair of endpoints associated with edge $e = (u, v)$. We define $\mathcal{C}_k \subset \mathcal{V}$ as the vertex set of the $k$-th cluster obtained via spectral clustering.

We identify the CBSG by its highly stable voltage profiles and robust electrical support. To identify nodes with strong static voltage-support capability, we define the node weights as follows:
\begin{equation}
	w_i = S_{sc,i} \cdot S_v^{-1} + \alpha\, C_i^{b},
\end{equation}
where $S_{sc,i}$ is the short-circuit capacity at node $i$, $S_v = \partial V_i / \partial Q_i$ is the voltage-reactive power sensitivity, $\alpha$ is a weighting coefficient, and $C_i^{b}$ is the betweenness centrality of node $i$, defined as the fraction of all shortest paths between every pair of distinct nodes that pass through $i$.

While these weights capture static voltage support, they do not reflect the grid's dynamic behavior under varying injections. We introduce a causal relationship matrix based on Granger causality:
\begin{equation}
	C_{ij}^{r} = C_{ij}^{G} \cdot
	\exp(-\lambda_d \cdot |Z_{ij}|),
\end{equation}
where $C_{ij}^{G}$ quantifies the predictive influence of node $i$ on node $j$ based on voltage time series and $\lambda_d$ is the electrical distance attenuation coefficient. We then derive the Net
Causal Strength (NCS) as:
\begin{equation}
	s_i^{cau} = \sum_j C_{ij}^{r} - \sum_j C_{ji}^{r},
\end{equation}
which distinguishes disturbance sources from sinks.

By integrating the node weights and the NCS, we identify $\mathcal{G}_{bone} = (\mathcal{V}_{bone}, \mathcal{E}_{bone})$ as a connected subgraph:
\begin{equation}
	\begin{array}{ll}
		\mathcal{V}_{bone} &= \{i \in \mathcal{V} \mid
		(w_i > \theta_w \lor s_i^{cau} > \theta_c)
		\land V_i^{base} \geq V_{th}\}, \\
		\mathcal{E}_{bone} &= \{(i,j) \in \mathcal{E} \mid
		i,j \in \mathcal{V}_{bone}\},
	\end{array}
	\label{equ:solve_CBSG}
\end{equation}
where $\theta_w, \theta_c$ are the weight and causality thresholds, $V_i^{base}$ denotes the nominal voltage level of bus $i$, and $V_{th}$ is the minimum voltage level threshold for backbone
membership.

With the CBSG identified, we divide the remaining power grid into multiple DHVCs whose behavior the backbone governs. To isolate local dynamics, we define the residual voltage conditioned on the backbone by subtracting the regression prediction from the non-backbone voltages:
\begin{equation}
	\tilde{\mathbf{V}} = \mathbf{V}_{\bar{B}}
	- \hat{\mathbf{V}}_{\bar{B}}, \quad
	\hat{\mathbf{V}}_{\bar{B}}
	= \boldsymbol{\Phi}\,\mathrm{PCA}_r(\mathbf{V}_B),
\end{equation}
where
$\mathbf{V}_B \in \mathbb{R}^{|\mathcal{V}_{bone}| \times N_s}$ and $\mathbf{V}_{\bar{B}} \in \mathbb{R}^{(|\mathcal{V}|-|\mathcal{V}_{bone}|) \times N_s}$ are the backbone and non-backbone voltage matrices over $N_s$ historical snapshots, $\mathrm{PCA}_r(\cdot)$ extracts the first $r$ principal components, and $\boldsymbol{\Phi} \in \mathbb{R}^{(|\mathcal{V}|-|\mathcal{V}_{bone}|) \times r}$ is a regularized regression coefficient matrix.

We construct the affinity matrix for clustering as:
\begin{equation}
	A_{ij} = \frac{1 + \rho(\tilde{V}_i, \tilde{V}_j)}{2}
	\cdot \exp\!\big(-\lambda_z\, Z_{ij}
	- \mu_t\, T_{ij}\big),
\end{equation}
where $\rho(\tilde{V}_i, \tilde{V}_j)$ denotes the Pearson correlation between the residual voltage sequences, $\lambda_z$ and $\mu_t$ are independent decay coefficients for the electrical impedance distance $Z_{ij}$ and the topological hop-count distance $T_{ij}$, respectively.

To ensure that the resulting clusters are both topologically connected and dynamically consistent, we partition the non-backbone vertices into $n_c$ clusters $\{\mathcal{C}_1, \ldots, \mathcal{C}_{n_c}\}$ via spectral clustering on the affinity matrix $A$:
\begin{equation}
	\begin{array}{cc}
		\min_{n_c} & \sum_{c=1}^{n_c}
		\left(1 - \dfrac{\lambda_1^{(c)}}{
			\|\tilde{\mathbf{V}}_{\mathcal{C}_c}\|_F^2}
		\right) \\[6pt]
		\mathrm{s.t.} & \text{each } \mathcal{C}_c
		\text{ is connected},
	\end{array}
	\label{equ:solve_DHVC}
\end{equation}
where $\tilde{\mathbf{V}}_{\mathcal{C}_c} \in \mathbb{R}^{|\mathcal{C}_c| \times N_s}$ is the residual voltage submatrix of cluster $\mathcal{C}_c$ and $\lambda_1^{(c)}$ is the leading eigenvalue of its covariance matrix. The objective penalizes deviations from a low-rank structure within each cluster.

Using a graph-theoretic approach, we decompose the CBSG and DHVCs into FTUs while preserving their electrical characteristics. Table~\ref{tab:graph_define} summarizes the definitions.

\begin{table*}[h]
	\centering
	\caption{Formal graph-theoretic definitions of fundamental
		circuit topologies}
	\label{tab:graph_define}
	\begin{tabular}{lcl}
		\hline
		\textbf{Type} & \textbf{Symbol}
		& \textbf{Mathematical Definition} \\
		\hline
		SERIES & $\mathbf{S}$
		& Path $P$ with $\deg_{\mathcal{G}}(v) = 2$
		for all interior vertices \\
		PARALLEL & $\mathbf{P}$
		& Edge set $\mathcal{E}_p \subseteq \mathcal{E}$
		where all $e \in \mathcal{E}_p$ share the same
		vertex pair \\
		RING & $\mathbf{R}$
		& Simple cycle with $\deg_{\mathcal{G}}(v) = 2$
		for every vertex on the cycle \\
		BRIDGE & $\mathbf{Br}$
		& Edge $e$ whose removal disconnects $\mathcal{G}$ \\
		MESH & $\mathbf{M}$
		& Irreducible subgraph that is neither series
		nor parallel \\
		TRANSFORMER & $\mathbf{T}$
		& Edge $e$ with tap ratio $\tau_e \neq 0$ \\
		TERMINAL & $\mathbf{L}$
		& Singleton cluster vertex $v$ with an inter-cluster
		edge \\
		INTER-CLUSTER & $\mathbf{I}$
		& Edge $e$ connecting $u \in \mathcal{C}_a$ and
		$v \in \mathcal{C}_b$, $a \neq b$ \\
		\hline
	\end{tabular}
\end{table*}

We first label branches to establish priority and exclusion criteria. We shield low-impedance edges during decomposition to prevent electrically nonsensical structures:
\begin{equation}
	\ell(e) = \begin{cases}
		\mathbf{T}  & \text{if } \tau_e \neq 0, \\
		\mathbf{SW} & \text{if } |z_e| < \epsilon_z, \\
		\bot        & \text{otherwise},
	\end{cases}
	\label{equ:transformer}
\end{equation}
where $\ell: \mathcal{E} \to \{\mathbf{T}, \mathbf{SW}, \bot\}$ is the initial edge labeling function, $\tau_e$ is the tap ratio of edge $e$, $z_e$ is its impedance, and $\epsilon_z$ is a small threshold. We use the label $\mathbf{SW}$ (switch) internally within the algorithm and exclude it from the final output.

We denote by $\mathcal{G}_R = (\mathcal{V}_R, \mathcal{E}_R)$ the residual graph after removing all edges labeled $\mathbf{T}$ or $\mathbf{SW}$, i.e., $\mathcal{E}_R = \{e \in \mathcal{E} \mid \ell(e) = \bot\}$. Through degree and cut-set analysis, we identify bridges and series chains:
\begin{equation}
	\begin{array}{ll}
		\mathbf{Br}(\mathcal{G}_R) &= \{e \in \mathcal{E}_R \mid
		\mathcal{G}_R \setminus \{e\}
		\text{ is disconnected}\}, \\[4pt]
		\mathbf{S}(\mathcal{G}_R) &= \bigcup_i P_i, \quad
		\text{where } \deg_{\mathcal{G}_R}(v) = 2,\;
		\forall v \in \operatorname{int}(P_i),
	\end{array}
\end{equation}
where each $P_i$ denotes a series chain in $\mathcal{G}_R$.

After we remove bridges and series chains, the residual graph decomposes into biconnected blocks. Let $\mathcal{G}_B = (\mathcal{V}_B, \mathcal{E}_B)$ denote one such block. We classify each block in descending order of structural specificity:
\begin{equation}
	\ell_B(\mathcal{G}_B) = \begin{cases}
		\mathbf{L}, & \text{if } |\mathcal{V}_B \cap \mathcal{C}_k|
		= 1 \text{ and } \exists\, e = (v,u) \in \mathcal{E}\\
		&\text{ with } v \in \mathcal{C}_k,\, u \notin
		\mathcal{C}_k, \\
		\mathbf{P}, & \text{if } \exists\, e_1, e_2
		\in \mathcal{E}_B,\, e_1 \neq e_2, \\&\text{ such that }
		\partial e_1 = \partial e_2, \\
		\mathbf{R}, & \text{if } \mathcal{G}_B
		\cong \mathrm{Cyc}_{|\mathcal{V}_B|} \text{ and }\\&
		\deg_{\mathcal{G}_B}(v) = 2,\,
		\forall v \in \mathcal{V}_B, \\
		\mathbf{S}, & \text{if } \deg_{\mathcal{G}_B}(v) = 2,\,
		\forall v \in \operatorname{int}(\mathcal{G}_B), \\
		\mathbf{M}, & \text{otherwise},
	\end{cases}
	\label{equ:ring_mash}
\end{equation}
where $\mathrm{Cyc}_n$ denotes the cycle graph on $n$ vertices.

By applying \eqref{equ:solve_CBSG} through \eqref{equ:ring_mash}, we transform the complex grid topology into a hierarchical tree of fundamental electrical structures. We then linearize each topological unit into:
\begin{equation}
	\mathrm{L}_{top} = \big\{\mathbf{Y}_{eq},\;
	\mathcal{P}_{ports},\; \mathcal{D}_{int}\big\},
\end{equation}
where $\mathbf{Y}_{eq} \in \mathbb{R}^{|\mathcal{P}_{ports}| \times |\mathcal{P}_{ports}|}$ is the equivalent admittance matrix of the unit, $\mathcal{P}_{ports} \subset \mathcal{V}$ is the set of boundary vertices connecting the unit to the rest of the network, and $\mathcal{D}_{int}$ contains the internal parameters required for recovering nodal quantities from port-level solutions.

This hierarchical decomposition replicates the control architecture of the power system at the data level, endowing the model with a physical inductive bias. When the topology changes, local disturbances mainly affect neighboring nodes while global regulation follows the power flow laws transmitted through the backbone network.

\subsection{Hierarchical bidirectional graph neural network}
\begin{figure}
	\centering
	\includegraphics[width=0.9\linewidth]{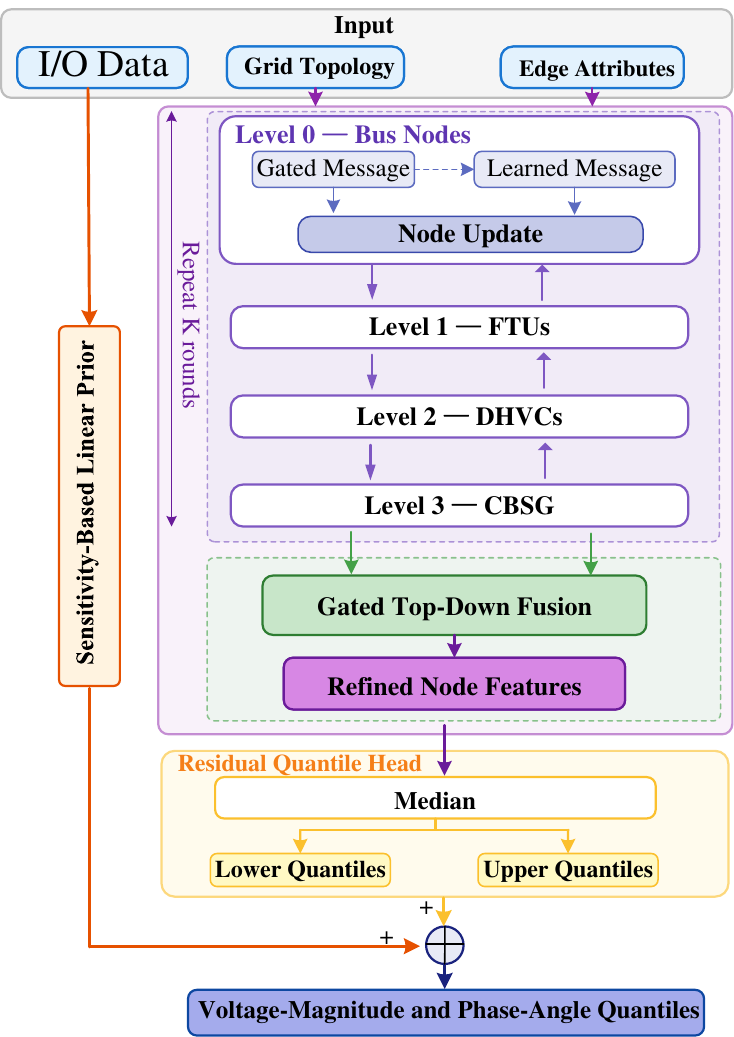}
	\caption{Architecture of HBGNN that illustrates the bidirectional hierarchical information exchange between Nodes, FTUs, DHVCs, and CBSG levels, and the quantile prediction module with linearization baseline.}
	\label{fig:placeholder}
\end{figure}

We design the hierarchical bidirectional graph neural network (HBGNN) to progressively compress the power grid into a four-level representation—nodes ($l=0$), FTUs ($l=1$), DHVCs ($l=2$), and CBSG ($l=3$)—to capture interactions at multiple spatial scales. The feature dimensions at each level are $\mathbf{h}^{(0)}\!\in\!\mathbb{R}^d$, $\mathbf{h}^{(1)}\!\in\!\mathbb{R}^d$, $\mathbf{h}^{(2)}\!\in\!\mathbb{R}^{2d}$, $\mathbf{h}^{(3)}\!\in\!\mathbb{R}^{4d}$. We perform a bidirectional message-passing process in which bottom-up aggregation extracts global patterns and top-down broadcasting refines local representations. The model predicts voltage magnitude and phase angle quantiles simultaneously through a shared HBGNN backbone with separate output heads:
\begin{equation}
	\begin{array}{ll}
		\hat{V}_\tau &= V_0 + S_{PV} \cdot \Delta P_\tau
		+ f_{HBGNN}^V(\Delta P, \mathcal{G}), \\
		\hat{\theta}_\tau &= \theta_0 + S_{P\theta} \cdot \Delta P_\tau
		+ f_{HBGNN}^\theta(\Delta P, \mathcal{G}),
	\end{array}
\end{equation}
where $V_0 \in \mathbb{R}^{b \times L_V}$ and $\theta_0 \in \mathbb{R}^{b \times L_\theta}$ are the base-case voltage magnitudes and phase angles at the operating point, $S_{PV} \in \mathbb{R}^{b \times q \times L_V}$ and $S_{P\theta} \in \mathbb{R}^{b \times q \times L_\theta}$ are sensitivity matrices that we fit via least squares in the normalized space, and $f_{HBGNN}^V(\cdot), f_{HBGNN}^\theta$ denote the HBGNN's nonlinear corrections. The architecture of HBGNN is shown in Fig. \ref{fig:placeholder}
\subsubsection{Bottom-up aggregation}
At the node level ($l=0$), we propagate messages along transmission edges via dual-pathway gated message passing with admittance weighting:
\begin{equation}
	\begin{array}{ll}
		\mathbf{G}_{ij} &= \sigma\!\big(\mathrm{MLP}_g(
		[\mathbf{h}_i \| \mathbf{h}_j \| \mathbf{e}_{ij}
		\| \epsilon_i])\big), \\[4pt]
		\mathbf{m}_{ij}^{g} &= \mathbf{G}_{ij} \odot
		\big[\sigma(\mathbf{W}_y Y_{eq,ij})
		\odot \mathbf{W}_m \mathbf{h}_j\big], \\[4pt]
		\mathbf{m}_{ij}^{ug} &= (\mathbf{1} - \mathbf{G}_{ij})
		\odot \mathrm{MLP}_m(\mathbf{h}_j), \\[4pt]
		\mathbf{m}_{ij} &= \mathbf{m}_{ij}^{g}
		+ \mathbf{m}_{ij}^{ug}, \\[4pt]
		\mathbf{h}_i^{(0)} &= \mathbf{W}_u
		\big[\mathbf{h}_i \| \textstyle\sum_{j \in \mathcal{N}_i}
		\mathbf{m}_{ij}\big].
	\end{array}
\end{equation}

We then pool features upward via three physics-guided operators—impedance-weighted attention for node-to-topology, $\epsilon$-aware attention for topology-to-cluster, and mean aggregation for cluster-to-backbone:
\begin{equation}
	\begin{array}{cl}
		\mathbf{h}^{(1)} &= \mathrm{MLP}^{(0\to1)}\!\Big(
		\sum_{i \in \mathcal{C}_p}
		\tfrac{e^{w_i / Z_i}}
		{\sum_{k' \in \mathcal{C}_p} e^{w_{k'} / Z_{k'}}}
		\mathbf{h}_i^{(0)}\Big), \\[6pt]
		\mathbf{h}^{(2)} &= \mathrm{MLP}^{(1\to2)}\!\Big(
		\sum_{j \in \mathcal{C}_c}
		\alpha_j\, \mathbf{h}_j^{(1)}\Big), \\[6pt]
		\mathbf{h}^{(3)} &= \mathrm{MLP}^{(2\to3)}\!\Big(
		\tfrac{1}{|\mathcal{C}_{bo}|}
		\sum_{c \in \mathcal{C}_{bo}}
		\mathbf{h}_c^{(2)}\Big), \\[6pt]
		\alpha_j &= \frac{
			\exp\!\big(\mathrm{MLP}_\alpha(
			[\mathbf{h}_j^{(1)} \| \epsilon_j^{(1)}])\big)}{
			\sum_{j' \in \mathcal{C}_c}
			\exp\!\big(\mathrm{MLP}_\alpha(
			[\mathbf{h}_{j'}^{(1)} \| \epsilon_{j'}^{(1)}])\big)},
	\end{array}
\end{equation}
where $w_i = \mathbf{W}_{attn}\mathbf{h}_i^{(0)} \in \mathbb{R}$ is the attention score, $Z_i \in \mathbb{R}$ is the impedance associated with the node-to-topology mapping, and $\epsilon_j^{(1)} = \sum_{i \in \mathcal{C}_j} \epsilon_i^{(0)} \in \mathbb{R}$ is the aggregated uncertainty at the topological level.

\subsubsection{Top-down broadcasting}
The global context is sent back via two complementary pathways. The hierarchical unpooling progressively maps features from level $l$ to $l{-}1$:
\begin{equation}
	\begin{array}{cl}
		\mathbf{h}_c^{(2)} &\leftarrow \mathbf{h}_c^{(2)}
		+ \mathrm{MLP}^{(3\to2)}(
		\mathbf{h}_{\pi^{(3)}(c)}^{(3)}), \\[4pt]
		\mathbf{h}_j^{(1)} &\leftarrow \mathbf{h}_j^{(1)}
		+ \mathrm{MLP}^{(2\to1)}(
		\mathbf{h}_{\pi^{(2)}(j)}^{(2)}), \\[4pt]
		\tilde{\mathbf{h}}_i^{hier} &=
		\mathrm{MLP}^{(1\to0)}(
		\mathbf{h}_{\pi^{(1)}(i)}^{(1)}).
	\end{array}
\end{equation}

In parallel, the mode-modulated broadcasting lets each node $i$ selectively aggregate backbone signals based on its mode affinity:
\begin{equation}
	\begin{array}{cl}
		\tilde{\mathbf{h}}_i^{mode} &=
		\sum_{bo \in \mathcal{V}^{(3)}} \beta_{bo,i} \cdot
		\big[\sigma(\mathbf{W}_{mode}\,\mathbf{z}_i)
		\odot \mathbf{W}_{bc}^{mode}\,
		\mathbf{h}_{bo}^{(3)}\big], \\[6pt]
		\beta_{bo,i} &= \frac{
			\exp(\mathbf{M}_{bo}^\top \mathbf{z}_i)}{
			\sum_{bo' \in \mathcal{V}^{(3)}}
			\exp(\mathbf{M}_{bo'}^\top \mathbf{z}_i)},
	\end{array}
\end{equation}
where $\mathbf{M}_{bo} = \mathbf{W}_{mp}\,\mathbf{h}_{bo}^{(3)} \in \mathbb{R}^M$ is the mode projection vector of backbone node $bo$, and $\mathbf{W}_{bc}^{mode} \in \mathbb{R}^{d \times 4d}$ projects backbone features to the node feature space.

We fuse the two pathways via a gating mechanism:
\begin{equation}
	\begin{array}{cl}
		\mathbf{h}_i^{(0)} &\leftarrow \mathbf{h}_i^{(0)}
		+ \boldsymbol{\gamma}_i \odot
		\mathrm{MLP}_{fuse}\!\big(
		[\tilde{\mathbf{h}}_i^{hier}
		\| \tilde{\mathbf{h}}_i^{mode}]\big), \\[4pt]
		\boldsymbol{\gamma}_i &= \sigma\!\big(\mathbf{W}_g\,
		[\mathbf{h}_i^{(0)}
		\| \tilde{\mathbf{h}}_i^{hier}
		\| \tilde{\mathbf{h}}_i^{mode}]\big).
	\end{array}
\end{equation}
After $K$ bidirectional rounds, each node obtains a refined feature vector $\mathbf{h}_i^{(0,K)} \in \mathbb{R}^d$ for quantile prediction.

\subsubsection{Quantile prediction and loss function}
To guarantee $\hat{V}_{i,\tau_1} \leq \cdots \leq \hat{V}_{i,\tau_Q}$, we apply a monotone quantile construction via cumulative softplus increments independently to both $V$ and $\theta$:
\begin{equation}
	\begin{array}{cl}
		\hat{y}_{i,cen} &= \mathrm{MLP}_{base}(
		\mathbf{h}_i^{(0,K)}), \\[4pt]
		\hat{y}_{i,cen \pm \tau} &= \hat{y}_{i,cen \pm (\tau-1)}
		\pm \mathrm{Softplus}\!\big(
		\mathrm{MLP}_\tau^\pm(\mathbf{h}_i^{(0,K)})\big),
	\end{array}
\end{equation}
where $\hat{y} \in \{\hat{V}, \hat{\theta}\}$. We combine the HBGNN residuals with the linearization baselines to obtain the final outputs: $\hat{\mathbf{V}}_i = \mathbf{V}_{lin,i} + [\hat{V}_{i,\tau_1}, \ldots, \hat{V}_{i,\tau_Q}] \in \mathbb{R}^{L_V}$ and $\hat{\boldsymbol{\theta}}_i = \boldsymbol{\theta}_{lin,i} + [\hat{\theta}_{i,\tau_1}, \ldots, \hat{\theta}_{i,\tau_R}] \in \mathbb{R}^{L_\theta}$.

We integrate the pinball losses with physics-informed constraints to form the total loss:
\begin{equation}
	\mathcal{L} = \sum_{y \in \{V, \theta\}}
	\big(\mathcal{L}_{pb}^{y}
	+ \lambda_{lin}\,\mathcal{L}_{lin}^{y}\big)
	+ \lambda_{kcl}(t)\,\mathcal{L}_{kcl}
	+ \lambda_{\theta}(t)\,\mathcal{L}_{\theta}^{sens},
\end{equation}
where $\mathcal{L}_{lin}^{y}$ penalizes deviation from the linearization baseline under small perturbations, $\mathcal{L}_{kcl}$ enforces Kirchhoff's current law via the admittance matrix, and
\begin{equation}
	\mathcal{L}_{\theta}^{sens} = \frac{1}{|\mathcal{I}|}
	\sum_{i \in \mathcal{I}}
	\mathcal{H}_\delta\!\big(
	\hat{\theta}_{i,med}
	- S_{P\theta,i}\,\Delta P_{med}\big)
\end{equation}
constrains phase angles to the power-angle sensitivity relation, where $\mathcal{I}$ denotes the set of injection buses, $\hat{\theta}_{i,med} \in \mathbb{R}$ is the median quantile prediction, and $S_{P\theta,i} \in \mathbb{R}^q$ is the $i$-th row of the sensitivity matrix. We set the warmup coefficients as $\lambda_{kcl}(t) = \lambda_{kcl}^0 \cdot \min(1,\, t / T_w)$ and $\lambda_{\theta}(t) = \lambda_{\theta}^0 \cdot \min(1,\, t / T_w)$, where $t$ is the training epoch and $T_w$ is the warmup period.

\section{Case study: Results and discussion}

\subsection{Evaluation design and statistical protocol}
\label{sec:experimental_setup}

The test systems are the IEEE 118-bus (bus\_118) and Polish 2746-bus (bus\_2746) systems. The PV inputs comprise 3{,}328 sets of 11-quantile forecasts from a PV power station at City University of Hong Kong during the summer of 2022~\cite{WANG2026114749}. PV penetration ranges from 10--95\% and 5--30\%, respectively. For each PV node, we draw $u\sim\mathcal{U}(0,1)$ and linearly interpolate the piecewise quantile function at $u$, yielding the forecast-implied marginal distribution without imposing a parametric form. Each forecast is used to generate 1{,}000 PV realizations, which are then evaluated using AC power flow, whose empirical bus-wise $V$ and $\theta$ quantiles form the targets. Loads and scheduled non-PV generation remain fixed at their base-case values. In our case study, an N--1 event removes one in-service branch, whereas an N--2 event removes two simultaneously. We would like to note that load uncertainty, generator redispatch, and cascading failures are outside the scope of this study.

After excluding nonconvergent contingencies, we split mutually exclusive outage events into train/validation/test sets at an $8{:}1{:}1$ ratio. The retained data contain 98/113 N--1/N--2 events for bus\_118 and 230/88 for bus\_2746. Cases are aggregated within events before event-level inference, preventing events with more retained cases from dominating the statistical summaries. All models share the partitions, targets, normalization, and evaluation code. Our experiments use Python~3.8.20 and PyTorch~2.4.1 on an i9-13900K workstation with 64~GB RAM and an RTX~4080~SUPER GPU.

Unless stated otherwise, learned models use hidden width 64 and AdamW with batch size 64, initial learning rate $10^{-3}$, weight decay $10^{-4}$, validation-based learning-rate reduction, and early stopping within 2{,}000 epochs. Table~\ref{tab:model_architectures} summarizes the evaluated methods.

\begin{table*}[!t]
	\caption{Configuration of the evaluated methods. All methods use separate monotone $V$/$\theta$ quantile heads.}
	\label{tab:model_architectures}
	\centering
	\footnotesize
	\setlength{\tabcolsep}{3pt}
	\renewcommand{\arraystretch}{1.04}
	\begin{tabular}{p{0.115\textwidth}p{0.17\textwidth}p{0.25\textwidth}p{0.36\textwidth}}
		\toprule
		Method & Model class & Main configuration & Structural/physical information \\
		\midrule
		TDD--HBGNN & Hierarchical GNN & 2 bidirectional rounds; 3 heads and 3 modes & Node--FTU--DHVC--CBSG hierarchy; impedance gating and sensitivity residual \\
		DNN~\cite{2024Bayesian} & Dense network & 4 layers; 256 internal units & Flattened PV input; no topology or physical prior \\
		GNN~\cite{lopezgarcia2025typedopf} & Edge-conditioned GNN & 6 layers; 64-dimensional states & Branch features $(r,x,|z|)$; no sensitivity prior \\
		PowerFlowNet~\cite{lin2024powerflownet} & TAGConv & 6 layers; polynomial order $K=3$ & Normalized high-order topology propagation \\
		GNS~\cite{donon2020neural} & Recurrent GNN & 6 message/update steps; GRU update & Branch-conditioned iterative propagation \\
		MD-GCN~\cite{gao2023physics} & Model-driven GCN & 6 layers; widths 64--256; 4 feature steps & Admittance-derived $G/B$ operators \\
		PG-GNN~\cite{yang2024ppf} & Chebyshev GNN & 3 layers; order $K=4$; widths 64--512 & Normalized graph-Laplacian propagation \\
		\bottomrule
	\end{tabular}
\end{table*}

The six baselines span dense, conventional graph, high-order, recurrent, model-driven, and spectral alternatives. They share the pinball objective and monotone heads, whereas TDD--HBGNN additionally uses linear-consistency and network-physics terms. Spatial diagnostics use MD-GCN for bus\_118 and DNN for bus\_2746 as stored global references. All-method analyses retain all six baselines.

Let $d^x_{eij}(\tau)=\hat Q^x_{eij}(\tau)-Q^{x,*}_{eij}(\tau)$ for output $x\in\{V,\theta\}$, event $e$ with $N_e$ cases, case $i$, bus $j$, and quantile $\tau$. For bus scope $\mathcal B_e$,
\begin{equation}
	\begin{aligned}
		D_1^x&=\frac{\sum_{i=1}^{N_e}\sum_{j\in\mathcal B_e}\int_{.05}^{.95}|d^x_{eij}(\tau)|\,\mathrm d\tau}{N_e|\mathcal B_e|(.95-.05)},\\
		M_\tau^x&=\frac{\sum_{i=1}^{N_e}\sum_{j\in\mathcal B_e}|d^x_{eij}(\tau)|}{N_e|\mathcal B_e|},\\
		T_{.05,.95}^x&=\frac{M_{.05}^x+M_{.95}^x}{2}.
	\end{aligned}
	\label{eq:distribution_metrics}
\end{equation}
We use trapezoidal integration on the nonuniform grids. Thus, $D_1^x$ measures average error across the stored distribution, whereas $T_{.05,.95}^x$ measures the two endpoint-quantile errors rather than empirical coverage. Topology-induced displacement is the analogous case-wise $D_1$ distance between the contingency and base-topology target functions under the same operating condition. The paired ratio $R_m(e)=m_e(\text{reference})/m_e(\text{TDD--HBGNN})$ favors TDD--HBGNN when it exceeds one. We report its median, win rate $\Pr[R_m(e)>1]$, and 95\% interval from 20{,}000 paired event-bootstrap replicates. Our ablations add one-sided paired Wilcoxon tests with Holm correction.

\subsection{Topology generalization and residual structure}
\label{sec:topology_overall}

\begin{figure}[!t]
	\centering
	\includegraphics[width=0.8\columnwidth]{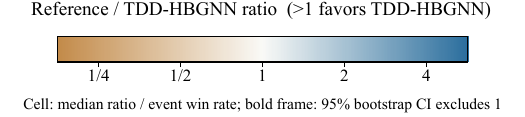}\\[-0.35ex]
	\subfloat[Within two graph hops.]{\includegraphics[width=0.485\columnwidth]{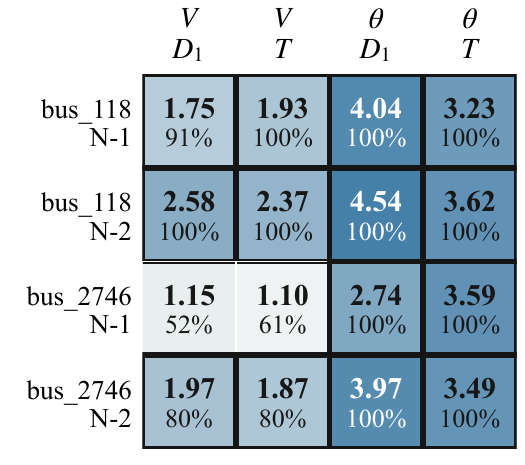}}
	\hfill
	\subfloat[All buses.]{\includegraphics[width=0.485\columnwidth]{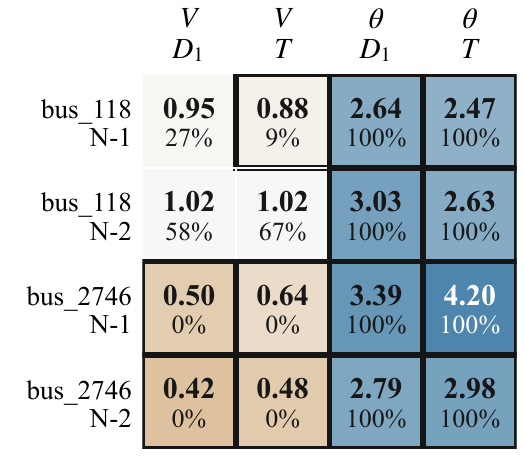}}
	\caption{Accuracy under unseen topologies. The cells provide the median reference/TDD--HBGNN ratio and TDD--HBGNN win rate, bold frames mark 95\% intervals excluding one. The references are MD-GCN for bus\_118 and DNN for bus\_2746.}
	\label{fig:topology_scope_matrices}
\end{figure}

Figure~\ref{fig:topology_scope_matrices} summarizes the results on unseen outage topologies within two graph hops of the outage and across the full network. The local scope tests sensitivity to changed power-transfer paths, whereas the global scope reveals whether local improvements persist after averaging over many weakly affected buses.

Near outages, the result is stable across metrics and contingency orders. On bus\_118, local $V$ ratios span 1.75--2.58 with 91--100\% event wins, while $\theta$ ratios span 3.23--4.54 with TDD--HBGNN winning every event. Under N--2, mean $D_1$ errors are $2.25\times10^{-5}$ versus $6.27\times10^{-5}$~p.u. for $V$ and $0.0734^\circ$ versus $0.757^\circ$ for $\theta$. On bus\_2746 N--2, the local $V$ and $\theta$ ratios are 1.97/1.87 and 3.97/3.49. The consistency between $D_1$ and endpoint errors indicates that the local advantage is not restricted to a single quantile.

The full-network panel defines the boundary. Angle ratios remain 2.47--4.20 and favor TDD--HBGNN in every event, but all-bus $V$ is near parity on bus\_118 and favors DNN on bus\_2746: $D_1/T_{.05,.95}$ ratios are 0.50/0.64 under N--1 and 0.42/0.48 under N--2. The local median ratio for bus\_2746 under N--1 is 1.15 with interval $[0.89,1.79]$, while its equal-event mean slightly favors DNN. Thus, the evidence supports system-wide angular and localized voltage accuracy, not universal all-bus $V$ superiority.

The mean--median reversal is driven by a few large penalties. For event 81--126, local $D_1^V$ is $3.82\times10^{-4}$ versus $2.55\times10^{-4}$~p.u. for TDD--HBGNN and DNN. Removing it reverses the mean ordering. The median therefore represents typical event ordering, whereas the mean remains sensitive to rare losses. Distance-decayed DHVC affinities concentrate learned $V$ corrections near disturbances, while the $S_{P\theta}\Delta P$ baseline and physics losses provide a broad angular response. This interpretation is tested by ablation, and spatial comparison alone does not establish causality.

\begin{figure}[!t]
	\centering
	\includegraphics[width=0.96\columnwidth]{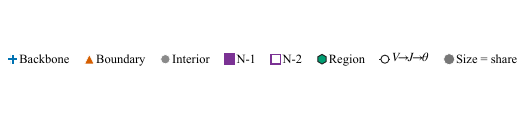}\\[-0.30ex]
	\subfloat[bus\_118.]{\includegraphics[width=0.485\columnwidth]{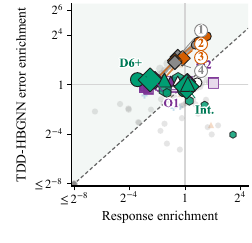}}
	\hfill
	\subfloat[bus\_2746.]{\includegraphics[width=0.485\columnwidth]{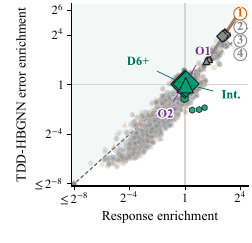}}
	\caption{Residual-error attribution over the complete eligible population. Log$_2$ axes show response and error enrichment, and marker area indicates Joint error share. Numbered bus markers and labeled regional markers identify leading contributors. Values below $2^{-8}$ lie on the displayed boundary.}
	\label{fig:spatial_error_attribution}
\end{figure}

We perform spatial attribution over the complete eligible population. Within each held-out event and output, bus-wise $D_1$ error and topology-induced displacement are normalized to unit mass. The Joint share is the average of the normalized $V$ and $\theta$ shares while Joint enrichment is their geometric mean. The analysis covers all 118/2{,}746 buses, all 11/12 and 23/10 held-out N--1/N--2 events, all 186/3{,}279 in-service physical branches, and all 15 predefined regional groups in each system. The 235 branches already out of service in the Polish base case have zero contribution. For each physical branch, endpoint-bus shares are divided equally among active incident branches, preserving unit mass. Figure~\ref{fig:spatial_error_attribution} retains the full bus and outage-event populations and labels only the leading contributors.

The exhaustive ranking reveals a localized hotspot in bus\_118. Buses 82, 83, and 96 and physical branches 82--83, 83--84, and 82--96 lead their respective rankings. The leading five and ten buses contribute 27.93\% and 42.47\% of Joint error, while the corresponding branch shares are 18.63\% and 30.63\%. The region at least six hops from the outages carries 54.24\% of the error, with error and response enrichments of 1.276 and 0.338. In contrast, bus\_2746 is more diffuse. Its leading five and ten bus shares are 5.73\% and 10.10\%, and its corresponding branch shares are 5.67\% and 9.72\%. The remote-region error and response enrichments are 1.013 and 0.970. Joint response--error correlations are 0.826/0.725 for bus\_118 and 0.950/0.947 for bus\_2746 under N--1/N--2. Thus, physical displacement closely tracks the spatial error allocation in the larger grid, whereas the smaller grid contains remote residual hotspots that displacement magnitude alone cannot explain.

\begin{figure}[!t]
	\centering
	\includegraphics[width=0.8\columnwidth]{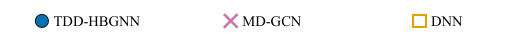}\\[-0.25ex]
	\subfloat[bus\_118.]{\includegraphics[width=0.485\columnwidth]{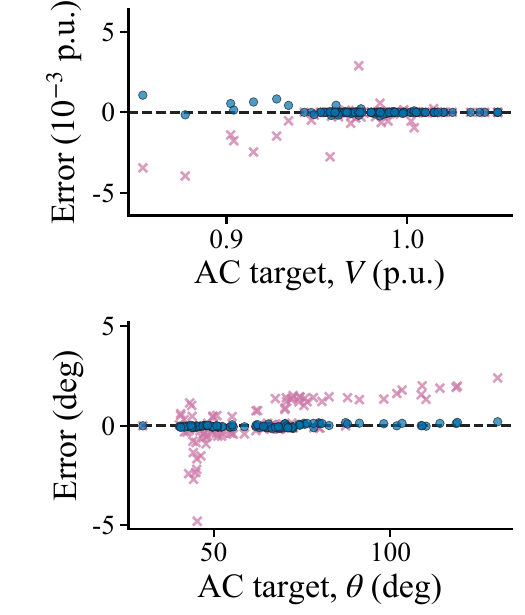}}
	\hfill
	\subfloat[bus\_2746.]{\includegraphics[width=0.485\columnwidth]{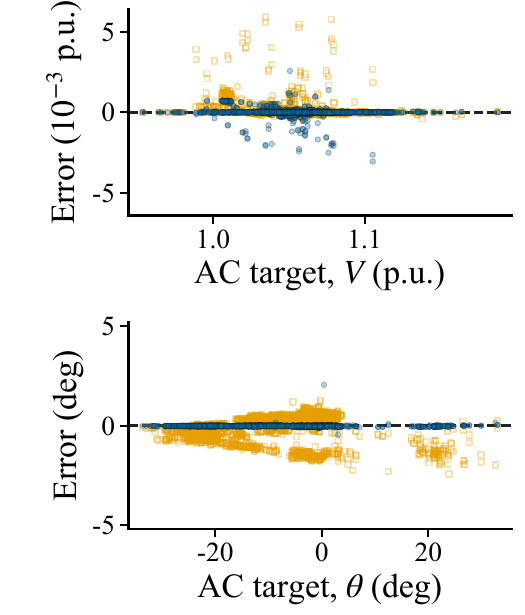}}
	\caption{Signed median-quantile bus errors in held-out cases.}
	\label{fig:pointwise_topology_cases}
\end{figure}

The results in Figure~\ref{fig:pointwise_topology_cases} indicate whether event averages conceal concentrated bus errors, while showing that all buses are retained in the pointwise comparison, and population results remain primary. For bus\_118, TDD--HBGNN reduces $V/\theta$ MAE from $2.46\times10^{-4}$~p.u./$0.704^\circ$ to $5.42\times10^{-5}$~p.u./$0.0619^\circ$. Although only 45.8\% of buses improve in $V$, the 95th-percentile and maximum errors fall by factors of 5.87 and 3.72. The $1/Z_i$ attention and uncertainty-aware pooling direct corrections toward strongly coupled residuals, reducing concentrated errors without requiring every bus to improve. On bus\_2746, 58.9\%/99.3\% of buses improve in $V$/$\theta$.

\subsection{Controlled scaling and all-baseline advantage}
\label{sec:controlled_reverse_scaling}

This experiment controls for physical response and operating conditions when comparing contingency orders. We verify disjoint event folders, use identical scenario indices, and pair N--1/N--2 events by minimum-cost matching on median and P95 topology-induced $V$/$\theta$ displacement, weighting the median twice. This is a minimum-discrepancy diagnostic, not equal-severity or causal matching. For matched pair $p$, let $\bar L_{k,x}^{(m)}(p)$ be method $m$'s mean case-level $D_1^x$ error over cases shared by order $k$. We define
\begin{equation}
	C_x^{(m)}(p)=
	\frac{\bar L_{\mathrm{N\text{--}2},x}^{(m)}(p)}
	{\bar L_{\mathrm{N\text{--}1},x}^{(m)}(p)},
	\qquad
	C_{\mathrm{joint}}^{(m)}(p)=
	\sqrt{C_V^{(m)}(p)C_\theta^{(m)}(p)}.
	\label{eq:joint_effect_ratio}
\end{equation}
Values above one indicate larger N--2 error. Joint is an equal-weight log-ratio summary rather than a training objective, so the $V$ and $\theta$ components remain visible.

\begin{figure}[!t]
	\centering
	\includegraphics[width=0.96\columnwidth]{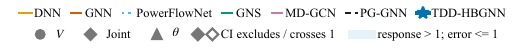}\\[-0.25ex]
	\subfloat[bus\_118.]{\includegraphics[width=0.485\columnwidth]{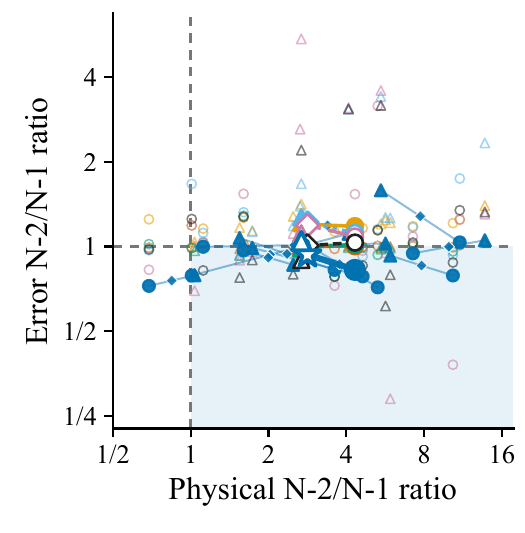}}
	\hfill
	\subfloat[bus\_2746 zoom.]{\includegraphics[width=0.485\columnwidth]{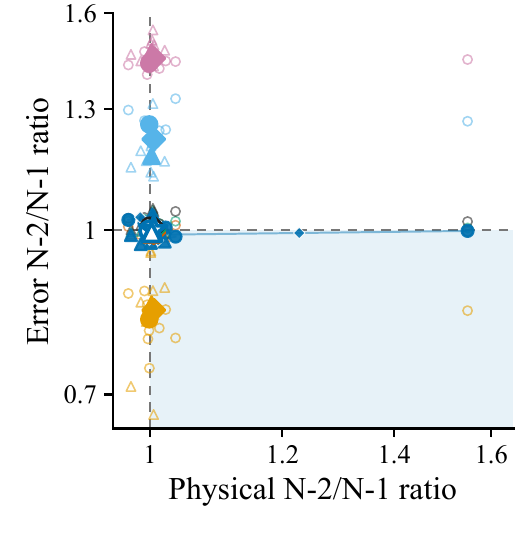}}
	\caption{Controlled physical-response and error ratios. Small symbols are matched events; large paths connect median $V$, Joint, and $\theta$. Open summaries have 95\% intervals crossing one.}
	\label{fig:controlled_reverse_scaling}
\end{figure}

Figure~\ref{fig:controlled_reverse_scaling} compares the matched physical-response ratios with the corresponding N--2/N--1 error ratios for both systems. On bus\_118, paired N--2 displacement rises by factors of 4.32 for $V$ and 2.68 for $\theta$, while TDD--HBGNN error ratios are 0.83/1.05/0.94 for $V$/$\theta$/Joint. The Joint interval $[0.86,1.01]$ includes parity, so easier pairs do not explain the lower N--2 point estimate. Both ratios are near parity on bus\_2746. A larger physical response therefore need not enlarge the residual when the reconstructed units, ports, and equivalent admittances remain compatible with the changed paths.

The results show that scaling behavior is method-dependent. DNN and PowerFlowNet errors increase on bus\_118, whereas MD-GCN and PowerFlowNet errors increase on bus\_2746. Several graph models remain near parity. TDD--HBGNN combines an admittance-weighted path with a learned path, retaining first-order network response while allocating nonlinear capacity to topology-dependent residuals. Contingency order is thus neither a direct severity scale nor a universal measure of learning difficulty.

\begin{figure}[!t]
	\centering
	\includegraphics[width=0.98\columnwidth]{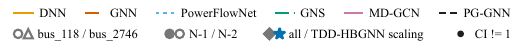}\\[-0.25ex]
	\subfloat[Controlled advantage.]{\includegraphics[width=0.485\columnwidth]{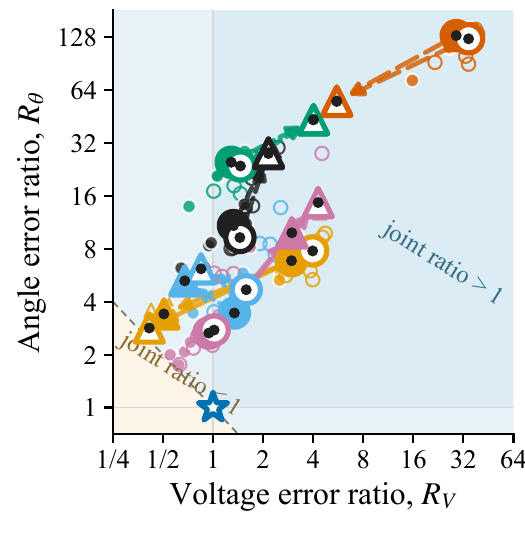}}
	\hfill
	\subfloat[Cross-system scaling.]{\includegraphics[width=0.485\columnwidth]{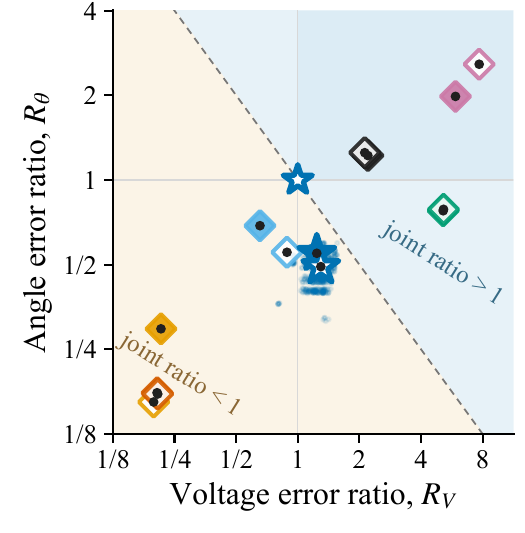}}
	\caption{Two-output ratio maps. Panel (a) uses event-median baseline/TDD--HBGNN ratios and panel (b) uses bus\_2746/bus\_118 ratios. Joint is the geometric mean of the $V$ and $\theta$ ratios. Black centers mark Joint 95\% intervals excluding one.}
	\label{fig:controlled_joint_advantage}
\end{figure}

All 24 summaries in Fig.~\ref{fig:controlled_joint_advantage}(a) lie above Joint parity with intervals excluding one. The smallest advantage, DNN on bus\_2746 N--2, combines a $V$ ratio of 0.41 with a $\theta$ ratio of 2.83, yielding Joint 1.09 $[1.06,1.13]$. On bus\_118, Joint ratios span 1.54--61.5 for N--1 and 1.68--60.2 for N--2. PowerFlowNet exhibits the same voltage--angle tradeoff on bus\_2746, but its Joint ratios remain 1.88/2.31. The sensitivity baseline and $\mathcal{L}_{\theta}^{sens}$ stabilize the broad angular response, whereas $V$ depends more strongly on the learned residual. These results support lower joint error, not component-wise dominance.

Fig.~\ref{fig:controlled_joint_advantage}(b) examines cross-system scaling at common penetration. For TDD--HBGNN, the bus\_2746/bus\_118 ratios are 1.24/0.55/0.82 under N--1 and 1.30/0.49/0.80 under N--2 for $V$/$\theta$/Joint. Across the evaluated methods, the Joint cross-system ratios fall below parity for DNN, GNN, and PowerFlowNet but exceed parity for GNS, MD-GCN, and PG-GNN. This change in direction across methods shows that the observed scaling behavior is method dependent. Although HBGNN reuses fixed-dimensional hierarchical operators across grids, results from only two systems cannot establish a universal relationship between grid size and prediction error.

\begin{figure}[!t]
	\centering
	\includegraphics[width=0.90\columnwidth]{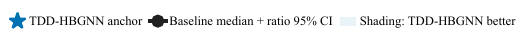}\\[-0.25ex]
	\subfloat[bus\_118: worst local 25\%.]{\includegraphics[width=0.485\columnwidth]{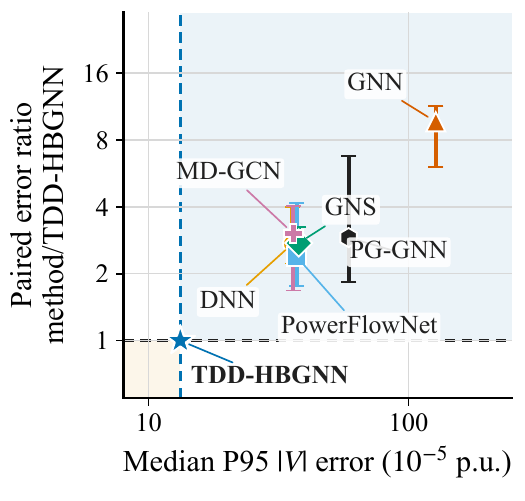}}
	\hfill
	\subfloat[bus\_2746: all local buses.]{\includegraphics[width=0.485\columnwidth]{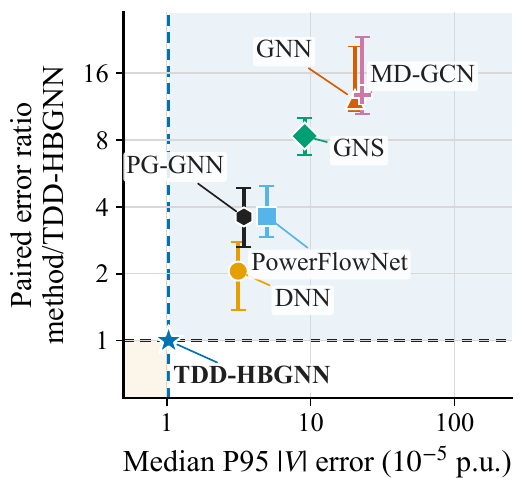}}
	\caption{Voltage-tail error at selected N--2/P95 coordinates. Axes show median event-level P95 error and paired baseline/TDD--HBGNN ratio, and whiskers are event-bootstrap 95\% intervals.}
	\label{fig:tail_all_method_dominance}
\end{figure}

A complementary test evaluates severe local conditions. Within two hops of an outage, each method's bus-wise $T_{.05,.95}^V$ errors are ranked per case, the worst retained fraction is averaged, and the P95 is taken across cases within each event. Figure~\ref{fig:tail_all_method_dominance} retains the worst 25\% of local buses for bus\_118 and all local buses for bus\_2746. It shows the median event-level P95 error and median paired baseline/TDD--HBGNN ratio at N--2 settings whose bootstrap lower bound exceeds one for every baseline.

TDD--HBGNN's median error is $13.25\times10^{-5}$~p.u. on bus\_118 versus $35.79$--$127.40\times10^{-5}$ for the baselines, and paired ratios span 2.39--9.66. On bus\_2746, it is $1.03\times10^{-5}$ versus $3.14$--$22.81\times10^{-5}$, with ratios 2.05--12.76. The linear term supplies first-order response, impedance pooling collects local residuals, and top-down broadcasting restores CBSG context.

\subsection{Finite-grid alignment and selective reliability}
\label{sec:alignment_boundary}

This experiment tests whether predicted quantiles reach their intended positions and whether a validation-trained score ranks high-error cases without test targets. We obtain the reached probability $\tilde\tau_{eij}(\tau)=(Q^{x,*}_{eij})^{-1}(\hat Q^x_{eij}(\tau))$ by piecewise-linear inversion with endpoint clipping. $\bar{\tilde\tau}_e(\tau)$ averages it over buses and cases within event $e$. The event descriptors, in percentage points, are
\begin{equation}
	\begin{aligned}
		E_{\mathrm{end}}&=\left|100[(\bar{\tilde\tau}_{.95}-\bar{\tilde\tau}_{.05})-.90]\right|,\\
		E_{\mathrm{int}}&=\max_{\tau\notin\{.05,.95\}}|100(\bar{\tilde\tau}(\tau)-\tau)|.
	\end{aligned}
	\label{eq:alignment_descriptors}
\end{equation}
$E_{\mathrm{end}}$ measures endpoint-mass error and $E_{\mathrm{int}}$ the largest interior displacement. These are finite-grid alignment diagnostics, not observation-based calibration measures.

\begin{figure}[!t]
	\centering
	\includegraphics[width=0.98\columnwidth]{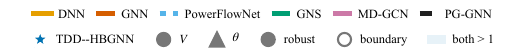}\\[-0.25ex]
	\subfloat[bus\_118, N--2.]{\includegraphics[width=0.485\columnwidth]{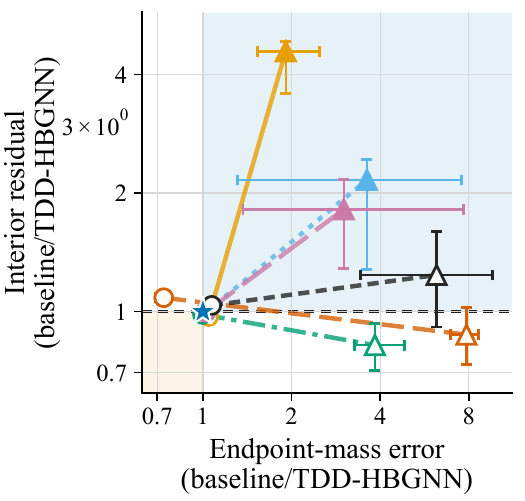}}
	\hfill
	\subfloat[bus\_2746, N--1.]{\includegraphics[width=0.485\columnwidth]{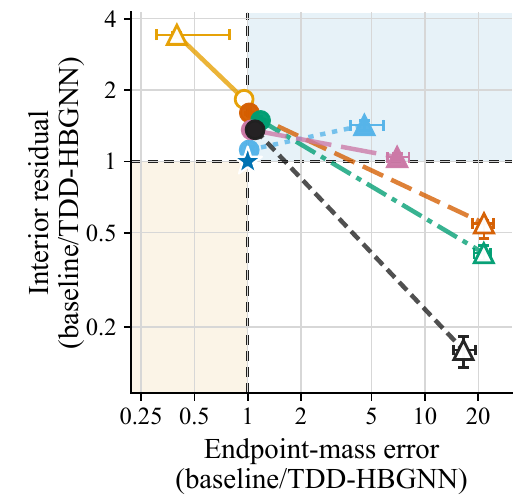}}
	\caption{Finite-grid alignment ratios. Axes are baseline/TDD--HBGNN endpoint and interior errors, and whiskers are event-bootstrap 95\% intervals. Circles/triangles denote $V$/$\theta$, and filled points have both intervals above one.}
	\label{fig:quantile_all_method_dominance}
\end{figure}

Figure~\ref{fig:quantile_all_method_dominance} reports all baseline--output coordinates for one selected contingency order per system. On bus\_118 N--2, 16 of 24 coordinate medians exceed one, but eight favor a baseline and no $V$ endpoint is robustly favorable on both descriptors. GNN and GNS trade endpoint ratios of 7.85 and 3.84 for interior ratios of 0.87 and 0.82. On bus\_2746 N--1, 19 coordinates robustly favor TDD--HBGNN, yet several reversals remain. TDD--HBGNN's central $V$ masses are only 43.77\%, 43.26\%, 72.59\%, and 72.54\%, far from 90\%. Its monotone head enforces quantile order, but the unconstrained sum of positive increments determines endpoint span. Pinball, linear-consistency, and network-physics losses do not directly enforce empirical coverage. Lower physical-unit error therefore does not establish voltage calibration.

The $\theta$ endpoints separate more consistently. On bus\_118, DNN, PowerFlowNet, and MD-GCN are worse on both descriptors, whereas GNN and GNS exchange endpoint error for smaller interior residual. On bus\_2746, DNN reaches 0.40/3.42, while GNN, GNS, and PG-GNN combine endpoint ratios 21.58/21.51/16.51 with interior ratios 0.55/0.41/0.16. The $S_{P\theta}\Delta P$ baseline and $\mathcal{L}_{\theta}^{sens}$ constrain angular response, but increment sums and their allocation govern endpoint span and interior shape separately. The observed reversals therefore represent distinct failure modes.

For selective reliability, a TDD--HBGNN pass yields corrected quantiles $q^x$ and a linear reference $q_{\mathrm{lin}}^x$. The physics risk head uses interval width, correction magnitude, corrected-to-linear width disagreement, curvature, and input deviation. A generic head omits the two physics-comparison features. Candidates are selected by event-grouped fivefold cross-validation on validation events and frozen before testing:
\begin{equation}
	s_\phi^x=h_\phi(\mathbf z^x),\qquad
	u^x=\widehat F_{\mathrm{val}}^x(s_\phi^x),\qquad C^x=1-u^x,
	\label{eq:risk_head}
\end{equation}
where $C^x$ is a rank index, not a correctness probability.

The realized tail error is the case-wise $T_{.05,.95}^x$ error averaged over buses. Specifically, Spearman's $\rho$ measures its monotone agreement with the risk score. nAUSE divides the area between score-based and oracle rejection curves by the random--oracle area. Zero denotes oracle, one random, and values above one worse-than-random ranking. Worst-decile lift measures score/error top-decile overlap relative to random overlap. The 50\% retained-error gain is $\log_{10}(E_{\mathrm{best,full}}/E_{\mathrm{kept},50})$, comparing retained TDD--HBGNN error with the best unscreened error among all seven methods. These metrics distinguish ranking quality from the absolute benefit of selective rejection.

\begin{figure}[!t]
	\centering
	\includegraphics[width=0.98\columnwidth]{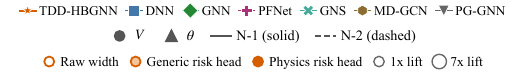}\\[-0.25ex]
	\subfloat[bus\_118.]{\includegraphics[width=0.485\columnwidth]{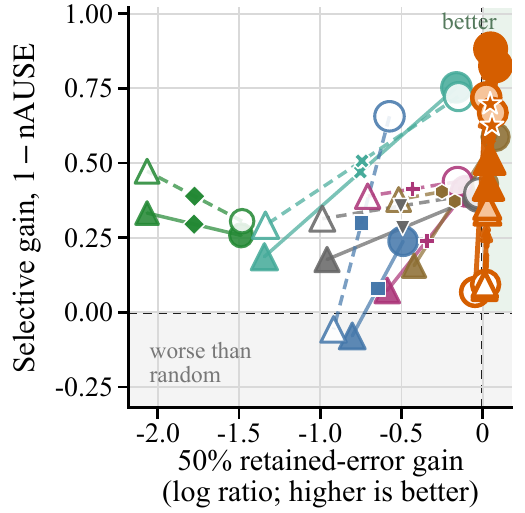}}
	\hfill
	\subfloat[bus\_2746.]{\includegraphics[width=0.485\columnwidth]{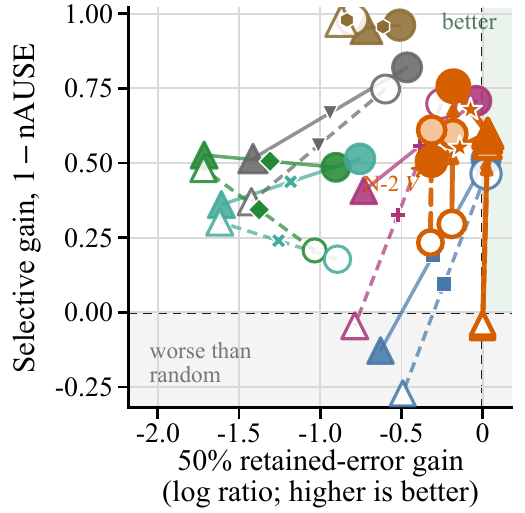}}
	\caption{Reliability--utility map. Vertical position is $1-\mathrm{nAUSE}$, and horizontal position is the 50\% retained-error gain. Marker area indicates worst-decile lift, and arrows progress from raw width to generic and physics risk heads.}
	\label{fig:confidence_reliability}
\end{figure}

Figure~\ref{fig:confidence_reliability} compares the reliability--utility trade-offs of raw interval width, the generic risk head, and the physics risk head. From the figure, raw width yields $\rho=0.0004$--0.206 and nAUSE $=0.70$--1.05. The physics head raises $\rho$ to 0.438--0.877, lowers nAUSE to 0.118--0.573, and gives 3.92--7.05-fold worst-decile lift. Width measures propagated uncertainty but not necessarily approximation error. Correction magnitude measures reliance on the nonlinear path; width disagreement detects departures from first-order spread; curvature detects uneven increments; and input deviation measures departure from validation support. A validation-fixed 90th-percentile score threshold flags 7.0--11.0\% of unseen cases and recalls 34.0--72.8\% of cases above the validation 90th-percentile tail-error threshold. Both $\rho$ and nAUSE improve in seven of eight settings.

For the bus\_2746 N--2 voltage exception, physics augmentation changes $\rho$ from 0.519 to 0.438 and nAUSE from 0.390 to 0.494, while lift changes from 3.70 to 3.65. A large nonlinear correction can be the correct response to topology redistribution. Treating it only as a risk indicator can therefore create false positives.

\subsection{Component evidence and deployment boundary}
\label{sec:component_ablation}

\begin{figure}[!t]
	\centering
	\includegraphics[width=0.94\columnwidth]{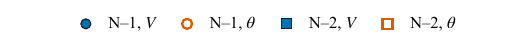}\\[-0.25ex]
	\subfloat[Within two graph hops.]{\includegraphics[width=0.485\columnwidth]{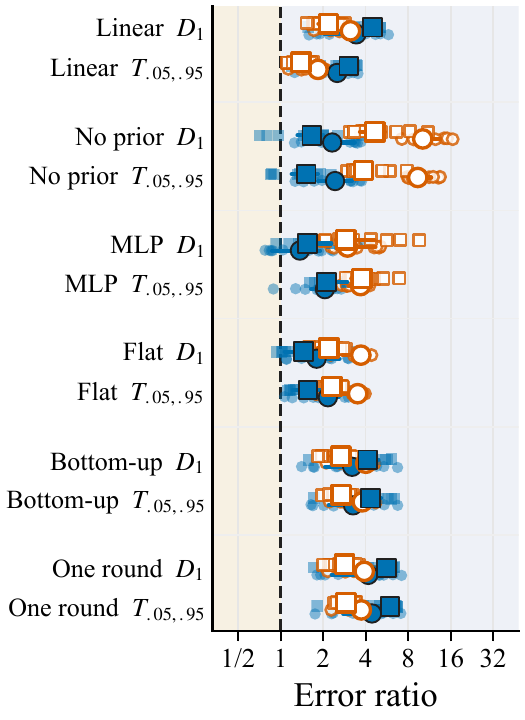}}
	\hfill
	\subfloat[All buses.]{\includegraphics[width=0.485\columnwidth]{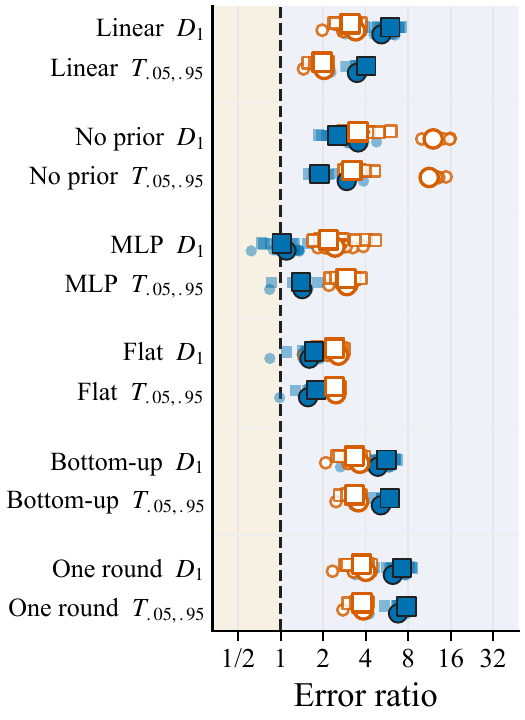}}
	\caption{Component ablation on bus\_118. Small symbols indicate events, large symbols and segments represent the median and IQR ($n=11/12$ for N--1/N--2). Ratios are calculated by variant/TDD--HBGNN, and tests are one-sided paired Wilcoxon with Holm correction.}
	\label{fig:component_scope}
\end{figure}

This section presents an ablation study to determine which components contribute to local and system-wide accuracy. We configure this test as follows. Linear only removes the learned correction, No linear prior removes the prior, MLP removes graph structure and the prior, Flat graph removes hierarchy, Bottom-up only removes top-down propagation, and One propagation round reduces two bidirectional rounds to one. All variants use the same held-out bus\_118 events and targets. Figure~\ref{fig:component_scope} reports accuracy effects, and Table~\ref{tab:computational_resources} reports computational profiles.

\begin{table*}[!t]
	\caption{Measured computational resources, training time, and non-ML wall-clock benchmarks.}
	\label{tab:computational_resources}
	\centering
	\footnotesize
	\setlength{\tabcolsep}{5.2pt}
	\renewcommand{\arraystretch}{1.02}
	\begin{tabular}{llcccc}
		\toprule
		Method or solver & Test system & Online wall time & Peak PyTorch memory & Parameters & Training time \\
		& & (ms/case) & (MB) & (million) & (h) \\
		\midrule
		\textbf{TDD--HBGNN} & bus\_118 & 0.163 / 0.161 & 94.1 / 94.9 & 0.658 / 0.658 & 2.220 / 0.665 \\
		& bus\_2746 & 1.793 / 1.770 & 1440.9 / 1403.7 & 0.658 / 0.658 & 2.46 / 2.96 \\
		\addlinespace[1pt]
		DNN & bus\_118 & 0.041 / 0.041 & 48.0 / 48.0 & 2.150 / 2.150 & 1.242 / 0.397 \\
		& bus\_2746 & 0.498 / 0.434 & 869.1 / 868.6 & 45.723 / 45.723 & 2.019 / 0.982 \\
		\addlinespace[1pt]
		GNN & bus\_118 & 0.069 / 0.067 & 52.3 / 52.2 & 0.194 / 0.194 & 0.937 / 0.946 \\
		& bus\_2746 & 0.952 / 0.887 & 749.7 / 747.0 & 0.194 / 0.194 & 2.836 / 0.752 \\
		\addlinespace[1pt]
		GNS & bus\_118 & 0.072 / 0.093 & 57.3 / 57.4 & 0.081 / 0.081 & 1.217 / 2.256 \\
		& bus\_2746 & 0.956 / 0.906 & 835.1 / 805.6 & 0.081 / 0.081 & 6.509 / 3.359 \\
		\addlinespace[1pt]
		MD-GCN & bus\_118 & 0.167 / 0.160 & 353.4 / 393.4 & 10.930 / 10.930 & 2.114 / 0.671 \\
		& bus\_2746 & 6.643 / 3.944 & 3513.8 / 3083.5 & 10.930 / 10.930 & 217.567 / 47.537 \\
		\addlinespace[1pt]
		PG-GNN & bus\_118 & 0.087 / 0.139 & 158.7 / 416.9 & 3.058 / 12.211 & 0.842 / 2.555 \\
		& bus\_2746 & 1.585 / 1.789 & 2422.3 / 2044.4 & 3.058 / 3.058 & 11.477 / 8.041 \\
		\addlinespace[1pt]
		PowerFlowNet & bus\_118 & 0.100 / 0.091 & 46.6 / 46.7 & 0.143 / 0.143 & 2.404 / 2.764 \\
		& bus\_2746 & 0.951 / 0.969 & 569.7 / 540.2 & 0.143 / 0.143 & 11.903 / 6.131 \\
		\midrule
		AC-PF (one solve) & bus\_118 & 15.095 / 15.108 & -- & -- & -- \\
		& bus\_2746 & 26.107 / 26.179 & -- & -- & -- \\
		MCS-PPF (1000 samples) & bus\_118 & 1{,}068 / 1{,}080 & -- & -- & -- \\
		& bus\_2746 & 1{,}890 / 1{,}914 & -- & -- & -- \\
		\bottomrule
	\end{tabular}

	\vspace{1pt}
	\begin{minipage}{0.97\textwidth}
		\scriptsize
    		\emph{Notes:} Entries are N--1 / N--2. Learned-model time is batch forward latency excluding data loading, host--device transfer, and post-processing. AC-PF entries are worker-measured mean solve times. MCS-PPF entries are end-to-end event times divided by 1000 cases, and each case uses 1000 AC-PF samples and 16 workers. Memory is process-local, and ``--'' means not applicable.
	\end{minipage}
\end{table*}

Removing either half of the prior--correction structure is costly. Linear only gives median $D_1$ ratios of 2.18--5.99, while No linear prior gives 1.66--12.04. The former retains the sensitivity response but cannot represent higher-order topology-dependent residuals. The latter forces the nonlinear network to learn both the dominant first-order mapping and its correction. Flat graph, Bottom-up only, and One propagation round yield 1.46--3.70, 2.67--5.93, and 2.84--7.80. We observe that all of them remain significant after Holm correction ($p\leq0.0264$). MLP's global $V$-$D_1$ ratios are 1.10 and 1.02 and are not significant, but its N--2 local ratio is 1.56, angle ratios exceed 2.18, and voltage-tail ratios exceed 1.39. The advantage therefore lies in angular, tail, and outage-neighborhood behavior rather than every all-bus voltage average.

The structural variants identify the required information path. Flattening removes node--FTU--DHVC--CBSG compression and weighted pooling, leaving node-level propagation to recover long-range dependence. Bottom-up-only propagation forms a system summary but cannot return it through hierarchical unpooling and broadcasting. One round broadcasts the initial summary once, whereas a second round re-aggregates globally corrected node states and closes the local--global feedback loop. Their degradation supports both multiscale compression and repeated bidirectional exchange.

Table~\ref{tab:computational_resources} shows that TDD--HBGNN retains 0.658 million parameters across both grids, whereas DNN grows from 2.150 to 45.723 million but remains faster in batch inference. The fixed parameter count follows from shared fixed-width hierarchical operators and does not imply constant training cost, because the larger grid requires more unit evaluations and activation memory. TDD--HBGNN training takes 2.220/0.665~h on bus\_118 and 2.46/2.96~h on bus\_2746. The variation reflects event counts, per-epoch cost, and early stopping. The MCS times correspond to 1{,}068/1{,}080 and 1{,}890/1{,}914~ms per case, making TDD--HBGNN $1.05\times10^3$--$6.71\times10^3$ times faster. Note that the one-solve AC-PF values represent only the deterministic MCS kernel, not an equivalent probabilistic output. These results support amortized screening, with high-risk cases routed to AC power flow or full probabilistic analysis.
\section{Conclusion}
\label{conclusion}
This paper developed TDD--HBGNN for PPF under changing grid topologies. The method separates transferable physical structure from topology-dependent nonlinear learning. For each topology, it reconstructs FTUs, DHVCs, and a CBSG and preserves electrical ports through equivalent-admittance representations. A sensitivity prior captures the dominant first-order voltage and angle response, leaving the HBGNN to learn the nonlinear residual. Impedance- and uncertainty-aware bottom-up pooling aggregates local perturbations, while mode-modulated top-down broadcasting returns system context to buses. Two exchange rounds close the local--global feedback loop. Shared fixed-width operators process the reconstructed hierarchy instead of flattened system-specific vectors, allowing the architecture to accommodate different grid sizes and within-system unseen contingencies.

Tests on the IEEE 118-bus and Polish 2746-bus systems validate these advantages. Under unseen N--1 and N--2 contingencies, TDD--HBGNN achieves all-bus phase-angle ratios of 2.47--4.20 against the strongest benchmarks. All 24 system--contingency--baseline summaries favor TDD--HBGNN in joint voltage--angle error. Our exhaustive attribution identifies a localized hotspot on the IEEE system, where the leading ten buses and physical branches contribute 42.47\% and 30.63\% of Joint error, and the remote region has error enrichment of 1.276 despite response enrichment of 0.338. The corresponding Polish contributions are 10.10\% and 9.72\%, while its remote-region error and response enrichments remain near parity. These associations localize error accumulation but do not establish causal branch effects. The ablation analysis shows that removing the prior or learned correction, flattening the hierarchy, suppressing top-down propagation, or using a single exchange round increases $D_1$ by median factors of 1.46--12.04, linking the gains to the prior--residual formulation and repeated hierarchical exchange.

Future work should extend the sensitivity prior to strongly nonlinear and unbalanced regimes and incorporate uncertainty from loads and non-PV dispatch. The quantile heads and validation-trained risk scores require topology-aware calibration, out-of-distribution detection, adaptive thresholds, and verified physical-solver fallback. Genuine zero-shot transfer should also be tested on related utility networks while keeping the learned hierarchical operators fixed.

\bibliographystyle{IEEEtran}
\bibliography{myref}

\end{document}